\documentclass{eptcs}
\providecommand{\event}{Refine 2011}

\usepackage{oz2e}

\newtheorem{definition}{Definition}

\def\A{{\bf A}}
\def\E{{\bf E}}
\def\X{{\bf X~}}

\newcommand{\Rel}[1]{{\mathsf{#1}}}

\title{Building a refinement checker for Z}
\author{John Derrick, Siobh\'an North and Anthony J.H. Simons
\institute{
Department of Computing, University of Sheffield,
Sheffield, S1 4DP, UK.}
\email{ J.Derrick@dcs.shef.ac.uk}}
\def\titlerunning{Building a refinement checker for Z}
\def\authorrunning{J. Derrick et al.}

\begin{document}



\maketitle

\begin{abstract}
In previous work we have described how refinements can be checked using a temporal logic based model-checker, and how
we have built a model-checker for Z by providing a translation of Z into the SAL input language. In this paper we
draw these two strands of work together and discuss how we have implemented refinement checking in our Z2SAL toolset.

The net effect of this work is that the SAL toolset can be used to check refinements between Z specifications supplied as
input files written in the \LaTeX mark-up. Two examples are used to illustrate the approach and compare it with a manual 
translation and refinement check. 
\end{abstract}

\section{Introduction}
\label{sec:introduction}

In this paper we  discuss the development of tool support for refinement checking in Z. In doing so we draw on two strands of work - one on providing a translation of Z into the input language of the SAL tool-suite, and the other on using model checking to verify refinements in state-based languages.

The SAL \cite{dem03a} tool-suite is used in both strands, and is designed to support the analysis and verification of
systems specified as state-transition systems. Its aim is to allow
different verification tools to be combined, all working on an input
language designed as a format into which programming and
specification languages can be translated.  The input language
provides a range of features to support this aim, such as guarded
commands, modules, definitions etc., and can, in fact, be used as a
specification language in its own right. The tool-suite currently
comprises a simulator and four model checkers  \cite{cla00a}  including those for LTL and CTL.

Our work on the first strand has resulted in a translation tool which converts Z specifications to 
a SAL module,
which groups together a number of definitions including types,
constants and modules for describing a state transition system.  
The declarations in a state schema in Z are translated into local variables in a 
SAL module, and any state predicates become appropriate invariants over the 
module and its transitions.

A SAL specification defines its behaviour by specifying transitions, thus it is 
natural to translate each Z operation into one branch of a guarded choice in the
transitions of the SAL module. The predicate in the operation schema
becomes a guard of the particular choice.
The guard is followed by a list of assignments, one for each output and primed
declaration in the operation schema. This methodology has been implemented in a tool-set, as described in \cite{DerrickNS06,DerrickNS08}.

Our work on the second strand has derived a methodology for verifying a refinement using a model-checker by combining two specifications in a special way and verifying particular CTL properties for this combination.
Specifically, \cite{SmithD05,SmithD06,DerrickS08} described how refinements in Z and other state-based languages could be verified by encoding downward and upward simulations as CTL theorems - the simulation conditions being the standard way to verify refinements in state-based languages such as Z, B etc.

The contribution we describe in this paper is to implement this methodology in our Z to SAL translation toolkit. This extension to the tool enables two Z specifications to be input in \LaTeX format, and for a refinement check to be performed. Internally this is achieved by translating each specification from \LaTeX to a single SAL specification upon which appropriate CTL theorems can be verified using the SAL CTL witness model-checker sal-wmc.

The purpose of this paper is to describe how this is done, using two examples as way of illustration. The structure of the paper is thus as follows.
In Section \ref{sec:refinement} and Section \ref{sec:overview} we provide background on refinement and the Z to SAL translation respectively. How specifications can be combined to enable a model checker to verify a refinement is described in Section \ref{sec:combination}, and this section also describes our implementation of this methodology. To illustrate the process we present a slightly more complicated example in Section \ref{sec:example} and we conclude in 
Section \ref{sec:conclusions}.

\section{Refinement}
\label{sec:refinement}

Data refinement \cite{DeRoever98,der01a} is a formal notion of development,
based around the idea that a concrete specification can be substituted
for an abstract one as long as its behaviour is consistent with that
defined in the abstract specification.

Each language, method or notation has its own variants. In Z, refinement is defined so
that the observable
behaviour of a specification is preserved. This behaviour is in
terms of the operations that are performed, and their input and output
values. Values of the state variables are regarded as being internal, and thus 
refinement can be used to change the representation of the state of a
system. Hence the term {\em data refinement\/}.

In a state-based setting such as provided by Z, data refinements are verified by
defining a relation (called a {\em retrieve relation\/}) between the
two specifications and verifying a set of {\em simulation conditions}. 
The retrieve relation shows how a state in one
specification is represented in the other. For refinement to be
complete, a relation, rather than simply a function, is required
\cite{der01a}.

In general, there are two forms the simulation conditions take, depending on the
interpretation given to an operation, specifically that given to the
operation's guard or precondition \cite{der01a}. The two
interpretations are often called the {\em blocking} and {\em
non-blocking} semantics. We consider the latter, i.e., the standard,
approach in this paper.

For any interpretation, there are two simulation rules for refinement which are together
complete, i.e., all possible refinements can be proved with a
combination of the rules. The first rule, referred to as {\em
downward\/} (or {\em forward\/}) {\em simulation\/} \cite{der01a,DeRoever98}, requires that
\begin{description}
\item[initialisation] the initial states of the concrete specification are related to
abstract initial states
\item[applicability] the concrete operations are enabled (at minimum) in states related to
abstract states where the corresponding abstract operations are
enabled, and
\item[correctness] the effect of each concrete operation is consistent with the
requirements of the corresponding abstract operation.
\end{description}

We do not consider the 
alternative kind of simulation known as an {\em upward} simulation in this paper, although there
is nothing to stop the the appropriate methodology being implemented in our tool suite.

\begin{definition}
\label{downsim}
A Z specification with state schema $CState$, initial state schema
$CInit$ and operations $COp_1 \ldots COp_n$ is a downward simulation
of a Z specification with state schema $AState$, initial state schema
$AInit$ and operations $AOp_1 \ldots AOp_n$, if there is a retrieve
relation $R$ such that the following hold for all $i : 1..n$.

\[1.~ \forall CState \dot CInit \implies (\exists AState \dot AInit 
\wedge R) \\
2.~ \forall AState; CState \dot R \wedge \pre AOp_i \implies \pre COp_i \\
3.~ \forall AState; CState; CState' \dot R \wedge \pre AOp_i \wedge COp_i \implies 
(\exists AState' \dot R' \wedge AOp_i  )\]
\end{definition}

The use of a retrieve relation allows the state spaces of the abstract and
concrete specifications to be different - the retrieve relation documents their
relationship. The first condition ensures appropriate initial states are
related, and the second that the concrete operations are defined whenever abstract ones are 
(modulo the retrieve relation). The third conditions ensures
that the concrete operations have an effect that is consistent with the
abstract, whilst also allowing non-determinism to be reduced. 

As an example refinement, consider the following simple specification. It defines two operations that add and remove an input from a set $s$ of some given type $T$.

\begin{sidebyside}
\[ [T] \]
\nextside
\begin{axdef}
max : \nat
\end{axdef}
\end{sidebyside}

\begin{sidebyside}
\[ A = [ s : \pset T \bbar \# s \leq max ] \]
\nextside
\[ AInit = [ A' \bbar  s' = \emptyset ] \]
\end{sidebyside}

\begin{sidebyside}
\begin{schema}{AEnter}
   \Delta A \\
   p? : T
\ST
  \# s < max \\
p? \nmem s \\
s' = s \cup \{ p? \}
\end{schema}
\nextside
\begin{schema}{ALeave}
   \Delta A \\
   p? : T
\ST
p? \mem s \\
s' = s \setminus \{ p? \}
\end{schema}
\end{sidebyside}

A simple data refinement replaces the set $s$ by an injective sequence $l$ as follows (assuming the same $T$ and $max$):

\begin{sidebyside}
\[C = [ l : iseq\ T \bbar \# l \leq max ]\]
\nextside
\[CInit = [ C' \bbar l' = \emptyseq]\]
\end{sidebyside}

\begin{sidebyside}
\begin{schema}{CEnter}
   \Delta C \\
   p? : T
\ST
  \# l < max \\
p? \nmem \ran l \\
l' = l \cat \lseq p? \rseq
\end{schema}
\nextside
\begin{schema}{CLeave}
   \Delta C \\
   p? : T
\ST
p? \mem \ran l \\
l' = l \filter (T \setminus \{ p? \})
\end{schema}
\end{sidebyside}

It is easy to see that the second specification is a downward simulation of the first, using as retrieve relation the following:

\[ 
R == [ A; C \bbar s = \ran l ]
\]
Our task is to build a tool that can automatically check this kind of refinement.

\section{Z2SAL}
\label{sec:overview}

The original idea of translating Z into SAL specifications was due to Smith and
Wildman \cite{smi05a},  however, our
implementation has increasingly diverged from the original idea as optimization
issues have been tackled. In \cite{DerrickNS06,DerrickNS08} we have described the basics of our 
implementation, which provides a bespoke parser and generator, written in
Java, to translate from the \LaTeX\  encoding of Z into the SAL input language. 

A Z specification written in the state-plus-operations style is translated into a SAL finite state automaton, 
following a template-driven strategy with a number of associated heuristics. The Z-style of specification is preserved in this strategy, including
postconditions that mix primed and unprimed variables arbitrarily, possibly 
asserting posterior states in non-constructive ways. We also preserve the Z
mathematical toolkit's approach to the modelling of relations, functions and
sequences as sets of tuples, permitting interchangeable views of functions,
sequences and relations as sets. 

A specification in the SAL input language consists of a collection of
separate input files, known as $contexts$, in which all the declarations are 
placed. At least one $context$ must contain the definition of a $module$,
an automaton to be simulated or checked. In our translation strategy, we use
a master $context$ for the main Z specification and refer to other $context$ 
files, which define the behaviour of data types from the mathematical toolkit.
The master $context$ consists of a prelude, declaring types and constants, 
followed by the main declaration of a SAL $module$, defining the finite
state automata, which implements the behaviour of the Z state and operation
schemas. The states of the SAL translation are created by aggregating the variables from the Z state schema,
and the transitions are created by turning the operation
schemas into $guarded\ commands$, triggered by preconditions on input and 
local (state) variables, and asserting postconditions on local and output
variables. 

The implementation of this basic strategy is presented in \cite{DerrickNS08}, here we recap on its salient points on two examples. Consider the first specification above. Upon translation the specification becomes a context, here called $a$. 

The {\em built-in} types of Z are translated into finite subranges in SAL, according to a scheme described in \cite{DerrickNS08}.
For example, $\nat$ is translated to:
\begin{verbatim}
NAT : TYPE = [0..4];
\end{verbatim}

The {\em basic types} of Z are converted into finite, enumerated sets in SAL, 
consisting of three symbolic ground elements by default (but sometimes 
with an extra $bottom$ element to deal with partiality of functions etc.). 
For example, the given type $T$ is translated to:
\begin{verbatim}
T : TYPE = {T__1, T__2, T__3};
\end{verbatim}




Where the Z specification expresses predicates involving the cardinality
of sets, the translator generates a bespoke counting-context for sets
containing up to the maximum number of symbolic ground elements generated 
for the set, as described in \cite{DerrickNS08}.  For this example, a
\verb+count3+ context is generated; the instantiation for counting up to
three elements of type \verb+T+ is named:
\begin{verbatim}
T__counter : CONTEXT = count3 {T; T__1, T__2, T__3};
\end{verbatim}

The bounding constant $max$ is an uninterpreted constant in Z, which we
translate in SAL as a local variable, which can in principle take any value
in the \verb+NAT+ type's range.  This leads to some simulation states where 
the limits of the system's behaviour are reached quickly (e.g. if $max$ = 0),
but other states in which all three elements may be added to the set \verb+s+.

{\bf State and initialisation schemas.} 
The state variables from the Z state schema are translated into the 
$local$ variables of the SAL $module$, which together constitute the
aggregate states of the automaton. The state predicate is treated as follows:
we define a corresponding \verb+DEFINITION+
clause to represent the schema invariant.  This is achieved by introducing an extra $local$ boolean variable, called
\verb+invariant__+, and declaring a formula for this in the $definition$ 
sub-clause.

The Z initialization schema is translated
in a non-constructive style into a guarded command in the 
\verb+INITIALIZATION+ clause of the SAL module, with the invariant as
part of the guard. Thus, for the above example, we get the following translation.

\begin{verbatim}
State : MODULE =
BEGIN                                                                 
 LOCAL max : NAT
 LOCAL s : set {T;} ! Set
 INPUT p? : T
 LOCAL invariant__ : BOOLEAN
 DEFINITION
   invariant__ = (T__counter ! size?(s) <= max)
 INITIALIZATION [
     s = set {T;} ! empty AND
     invariant__
   -->                                                                   
 ]
\end{verbatim}

The challenge of the translation strategy is to deal efficiently with the large vocabulary of mathematical data types such as sets, products, 
relations, functions, sequences and bags.  The translation tool has to represent these efficiently in SAL, whilst
preserving the expressiveness and flexibility of the Z language.

The basic approach is to define one or more context files for each data type in 
the toolkit.  
For example, the set mathematical data type in Z is translated into a
SAL context, which models the set as a boolean-valued membership predicate
on elements (following Bryant's optimal encoding of sets for translation
into BDDs, \cite{Bryant86,Bryant92}).  All other set operations are based 
on this encoding:

{\small
\begin{verbatim}
set {T : TYPE; } : CONTEXT = BEGIN
Set : TYPE = [T -> BOOLEAN];
empty : Set = LAMBDA (elem : T) : FALSE;
...
contains? (set : Set, elem : T) : BOOLEAN = 
 set(elem);
...
union(setA : Set, setB : Set) : Set =
 LAMBDA (elem : T) : setA(elem) OR setB(elem);
...
END
\end{verbatim}
}
Similar contexts are defined for the function, relation and sequence data
types.  Whereas Z sets and relations are modelled as boolean maps, Z functions
and sequences are modelled using SAL's total functions.  We adopt a totalising
strategy, introducing bottom elements for types that participate in the
domain or range of functions, or range of sequences.

{\bf Translating the Z operation schemas.}
Each operation schema in Z contributes in two ways to the SAL translation.
Firstly, an operation schema may optionally declare input, or output 
variables (or both), which are extracted and declared in the prelude 
of the $module$ clause, as SAL $input$ and $output$ variables. 
Secondly, the predicate of each operation schema is converted into a 
$guarded\ command$ in the $transition$ sub-clause, the last sub-clause
in the $module$ clause.

The input and output variables are understood to exist in the local scope 
of each operation schema, which has consequences in the translation.
The SAL translation eventually substitutes the suffix 
`\verb+_+'`\verb+_+' for `\verb+!+'
in the output variables, since the latter is reserved. 

The computation performed by each operation schema is expressed as a
$guarded\ command$ in the $transition$ sub-clause. The name of the schema
is used for the transition label, which aids readability. The $guarded\
command$ has the general syntactic form: \verb+label : guard --> assignments+.

The guards for each transition include the primed \verb+invariant__'+ as one
of the conjuncts, which asserts the state predicate in the posterior
state of every transition. This, combined with the assertion of the
unprimed \verb+invariant__+ in the initial state, ensures that the state
predicate holds universally.

Finally, a catch-all \verb+ELSE+ 
branch is added to the guarded commands, to ensure that the transition 
relation is total (for soundness of the model checking).  In practice, 
this allows model-checking to complete, even if the simulation blocks
at a given point.  Admitting the \verb+ELSE+-transition
allows simulations to pass through states in which the \verb+invariant__'+
fails to hold.  Normally, this does not matter, since we can also ensure
that \verb+LOCAL+ state variables are not modified, whenever the 
\verb+ELSE+-transition is taken.  

However, a new soundness problem emerged when admitting $bottom$ values, as
part of a totalising strategy for partial types.  Our previous practice was 
to assert that \verb+INPUT+ variables never took $bottom$ values, as part
of the invariant.  However, a loophole was discovered that allowed
the system to pass through states in which the invariant did not 
hold (due to selecting bottom values for inputs) and then recover in the 
following cycle, in which the invariant held once more, but undefined values
had been accepted as inputs from the previous cycle.  Ideally, we would have 
liked to rule out invalid inputs in the \verb+ELSE+-transition, but the SAL
tools do not permit this.

Instead, we now assert both the primed \verb+invariant__'+ and unprimed
\verb+invariant__+ in the guard to each transition, so closing the loophole.
In practice, simulations can still pass through states where the invariant 
fails to hold, but they are then forced to pass through \verb+ELSE+-transitions
repeatedly, until some valid input is selected.  The new translation is once 
again sound, but simulations may have more latent cycles.  Thus for the transition 
component of our example we have the following:

\begin{verbatim}
 TRANSITION [
   AEnter :
       T__counter ! size?(s) < max AND
       NOT set {T;} ! contains?(s, p?) AND
       s' = set {T;} ! insert(s, p?) AND
       invariant__ AND
       invariant__'
     -->
       s' IN {x : set {T;} ! Set | TRUE}
   [] 
   ALeave :
       set {T;} ! contains?(s, p?) AND
       s' = set {T;} ! remove(s, p?) AND
       invariant__ AND
       invariant__'
     -->
       s' IN {x : set {T;} ! Set | TRUE}
   []
   ELSE -->  s' = s
 ]                 
\end{verbatim}

A similar translation is produced for $C$, this time producing a SAL input file using 
contexts defined to model Z sequences; see Appendix A.

\section{Model-checking a refinement}
\label{sec:combination}

A series of approaches to model-checking a refinement is described in 
\cite{SmithD05,SmithD06,DerrickS08} by Smith and Derrick with varying degrees of 
sophistication. They all work by taking two specifications, $A$ and $C$ say, and building 
a combined system $M$ which encodes the behaviour of both in such a way that it is possible 
to write CTL properties to check the various aspects that are needed for simulation 
conditions to hold.
There are variations to this approach as follows.

\begin{enumerate}
\item 
Three different combinations are formed, $M_{init}$, $M_{app}$, $M_{corr}$, one for each of the three downward simulation conditions (and a similar methodology for upward simulations);
\item
One combination is formed, $M$, encoding all three properties to be checked in one system.
\end{enumerate}

These two approaches need the candidate retrieve relation to be passed to the tool, thus a final approach is
\begin{itemize}
\item Additionally have the model-checker search to find if such a retrieve relation exists.
\end{itemize}

For efficiency reasons (and here to aid readability) we describe our implementation of the first approach, again restricting ourselves for brevity to downward simulations. Thus in the approach we describe, which is an abbreviated discussion of \cite{SmithD06}, here three systems are formed and if all three checks are satisfied then the concrete system is indeed a downward simulation of the abstract system with the chosen retrieve relation.

To illustrate the approach, we use the example specified above, noting that although for readability we describe it as a combination of Z schemas, in our implementation the combination acts at the level of combining SAL modules. We will combine the two specifications into one system so that we can encode the simulation conditions on the combined system, thus the combined specification includes all the abstract and concrete variables. 
The methodology assumes the state variables of the abstract and concrete
systems are disjoint (as in fact they are in our example), but if not, then renaming is applied first to achieve it.

\medskip
{\bf Initialisation.} 
The simulation condition on initial states
requires that for each concrete initial state, we are able to find an
abstract initial state related by the retrieve relation $R$. To encode this condition
we initialise $M_{init}$  so that the concrete part of the state is initialised. Hence in our
example, the combined system's state and initialisation are as follows:

\begin{sidebyside}
\begin{schema}{M_{init}}
s : \pset T \\
l : iseq\ T
\ST
\# s \leq max \\
\# l \leq max
\end{schema}
\nextside
\begin{schema}{Init_{init}}
M_{init}'
\ST
   l' = \emptyseq
\end{schema}
\end{sidebyside}

To check whether an abstract initial state exists that is related to any particuar concrete
initial state, we use just one operation (normally called $InitA_{init}$) which changes
the abstract part of the state to an initial value and leaves the
concrete part unchanged. In our example this operation is then:

\[ [\Delta M_{init} \bbar s' = \emptyset \land l'=l]\]

For any non-trivial specification $InitA_{init}$ is total, thus we do not need the "catch-all" ELSE branch
in the SAL model-checker which is needed for non-total systems as described above.
%
Then, with a system with one operation the required initialisation condition holds if the
operation can be performed such that the resulting abstract and
concrete parts of the state are related by $R$. That is, we require that there exists a next state such that $s = \ran l$, i.e.:

\[\E\X (s = \ran l) \]


\medskip
{\bf Applicability.} 
Applicability conditions in refinements check the consistency of the operations' preconditions. To encode this as a temporal formula we introduce a variable $ev$ to the combined state to denote the name of the last
operation that occurred, and, as in \cite{SmithD06}, we use a different $\Rel{font}$ for the values of
type $ev$. Since we will need an additional operation to ensure totality, the combined state for an applicability check in our example will be the following:

\begin{schema}{M_{app}}
s : \pset T \\
l : iseq\ T \\
ev:\{\Rel{AEnter}, \Rel{CEnter}, \Rel{ALeave}, \Rel{CLeave}, \Rel{Choose}\}
\ST
\# s \leq max \\
\# l \leq max 
\end{schema}

The applicability condition requires that if abstract and concrete states are related by the retrieve relation, then the concrete operation must be applicable whenever the abstract one was. For the sake of efficiency we initialise to states which are already related by the retrieve relation, that is, here of the form\footnote{The value of $ev$ can be left underspecified.}:

\[
Init_{app} = [ M_{app}' \bbar s' = \ran l' ]
\]

Operations are then specified, one for each abstract or concrete operation, each shadowing the behaviour of the original operation, and only specifying the values of that operation (eg $AEnter_{app}$ defines values for variables that originate from the abstract specification). In addition, we introduce a $Choose$ operation.

\begin{sidebyside}
\begin{schema}{AEnter_{app}}
   \Delta M_{app} \\
   p? : T
\ST
  \# s < max \\
p? \nmem s \\
s' = s \cup \{ p? \} \\
ev'=\Rel{AEnter}
\end{schema}
\nextside
\begin{schema}{ALeave_{app}}
   \Delta M_{app} \\
   p? : T
\ST
p? \mem s \\
s' = s \setminus \{ p? \} \\
ev'=\Rel{ALeave}
\end{schema}
\end{sidebyside}

\begin{sidebyside}
\begin{schema}{CEnter_{app}}
   \Delta M_{app} \\
   p? : T
\ST
  \# l < max \\
p? \nmem \ran l \\
l' = l \cat \lseq p? \rseq\\
ev'=\Rel{CEnter}
\end{schema}
\nextside
\begin{schema}{CLeave_{app}}
   \Delta M_{app} \\
   p? : T
\ST
p? \mem \ran l \\
l' = l \filter (T \setminus \{ p? \})\\
ev'=\Rel{CLeave}
\end{schema}
\end{sidebyside}

\[Choose_{app} \sdef [\Delta M_{app}| ev'=\Rel{Choose}]\]

The applicability check can now be written in CTL as
follows.

\[
(\E\X(ev=\Rel{AEnter}) \implies \E\X(ev=\Rel{CEnter})) \wedge (\E\X(ev=\Rel{ALeave}) \implies \E\X(ev=\Rel{CLeave})) 
\]

%

{\bf Correctness.}
A similar methodology is applied to check the correctness condition, and here we use the same combined state and initialisation as used for applicability, as well as the same totalisation $Choose$:

\[M_{corr} \sdef M_{app}\\
Init_{corr} \sdef Init_{app}\\
Choose_{corr} \sdef Choose_{app}\]

The downward simulation correctness condition requires that any after-state of a concrete operation is related by the retrieve relation to an after-state of the abstract operation. To encode this correctly one needs to ensure that each operation in the combined state does not alter variables from the portion of state it is not representing. Thus we have operations of the form:

\[AOp_{corr} \sdef [AOp_{app}| l'=l]\\
COp_{corr} \sdef [COp_{app}| s'=s]\] 

This allows us to perform the operations $COp_{corr}$ and $AOp_{corr}$ in
sequence so that the abstract part of the final state reached is
identical to that which could have been reached by performing only
$AOp_{corr}$, and the concrete part is identical to that which could have
been reached by performing only $COp_{corr}$. The correctness condition is then:

\[
\E\X(ev=\Rel{AEnter}) \implies \A\X(ev=\Rel{CEnter} \implies \E\X(ev=\Rel{AEnter}
\land R)) \\
\qquad \wedge \\
\E\X(ev=\Rel{ALeave}) \implies \A\X(ev=\Rel{CLeave} \implies \E\X(ev=\Rel{ALeave}
\land R))
\]


{\bf Implementation in SAL.} The above is described in terms of combinations of Z specifications, although, of course, it is implemented in terms of combining SAL modules in our tool-suite. 

The process of combining the two \LaTeX\ Z specifications plus retrieve relation into a single SAL specification in order to check the downward simulation conditions
was achieved using an extension to our Z to SAL parser. When translating a single Z specification to
SAL our compiler first parses the Z, then transforms it into an internal SAL representation and finally
the SAL file is generated. In extending the tool-set to combine two specifications in the manner described above the major modification was
to the middle phase, the transformation from Z to SAL. Nevertheless the process of parsing two
specifications sequentially required some modification for a number of issues.

For example, declarations in the abstract and concrete state schemas need to be checked to ensure that they contain distinct identifiers, 
but where types and constants occur in both specifications they have to be
identical to cope with SAL's strict type checking. Neither of these problems caused much difficulty since, e.g.,
there already was a mechanism to ensure that a variable name used in two different Z operations
did not lead to a conflict in the SAL translations (where all variables had the same scope). In our
simple, single specification, translation this is achieved by prefixing the variable name by the name
of its transition wherever an ambiguous name is detected and the same mechanism was used when
producing a single combined specification. The only modification was that variables from axiomatic definitions were prefixed by the specification name rather than the transition
name.

Treating types declared in two different specifications as the same was slightly more complicated as
types from the abstract specification occurring in the concrete had to be identified. In our single
translation types are canonical, for reasons explained in \cite{DerrickNS08} and this had to be maintained in the combined
translation without the parser rejecting a concrete specification which contains an apparently
second declaration of a type which has been declared in the abstract specification. This problem
also occurred with identical constants in both specifications.

Having parsed the two specifications, the retrieve relation is read in and parsed as a single Z
operation with everything from both the abstract and concrete specifications in scope.

The process of transforming a single Z specification into SAL consists of fixing the finite ranges of all
the types, eliminating redundant predicates, giving initial values to all the constants and identifying
any named types that would have to be generated in SAL.  In
transforming two specifications into one SAL specification the finite ranges were fixed to the widest
required by either specification but apart from that the process is essentially simple. The two sets of
initial declarations were combined and the two lists of operation schemas in Z became a single list of
transitions in SAL. 
The resulting structure is that of our internal representation of any SAL specification and
a SAL text file could be generated from it in the standard way.

The result produced by our tool-kit of  the two SAL modules for the correctness condition is given in Appendix B. It is then a trivial matter to check the required theorem on it.

\section{A further example}
\label{sec:example}

A further example, which provides a comparative analysis with the manual approach to refinement checking, is given by the following (now standard) example.

The Marlowe box office allows customers to book tickets in advance using the $Book$ operation -- $mpool$ is the set of tickets, and if a ticket is available ($mpool\neq\emptyset$) then one
is allocated then and there. When the customer arrives, operation $Arrive$ presents this ticket. $Ticket$ is the set of all tickets, and a
free type adds a possibly null ticket, and $tkt$ models which tickets have been allocated.

\begin{sidebyside}
\[ [Ticket] \]
\nextside
\[ MTicket ::= null \bbar ticket \lang Ticket \rang \]
\end{sidebyside}

{\small
\begin{sidebyside}
\begin{schema}{Marlowe}
mpool:\pset Ticket\\
tkt: MTicket
\end{schema}
\nextside
\begin{schema}{MInit}
Marlowe
\ST
tkt = null
\end{schema}
\end{sidebyside}

\begin{sidebyside}
\begin{schema}{MBook}
\Delta Marlowe
\ST
tkt = null\\
mpool\neq\emptyset\\
tkt' \neq null\\
ticket^{-1}(tkt') \in mpool \\
mpool' = mpool \setminus \{ticket^{-1}(tkt') \}
\end{schema}
\nextside
\begin{schema}{MArrive}
\Delta Marlowe\\
t!:Ticket
\ST
tkt \neq null\\
tkt' = null\\
t! = ticket^{-1}(tkt)\\
mpool' = mpool
\end{schema}
\end{sidebyside}
}
In an alternative description - the Kurbel - customers still book tickets in advance. However, now if there is an available ticket then this is simply recorded by the operation $Book$ provided the customer has not already booked.
Only when the customer actually arrives at the box office, is the ticket allocated by $Arrive$. $kpool$ is the pool of tickets and $bkd$ denotes whether a ticket has been booked. 

{\small
\begin{sidebyside}
\[ Booked ::= yes \bbar no \]
\nextside
\[ [ Ticket] \]
\end{sidebyside}

\begin{sidebyside}
\begin{schema}{Kurbel}
kpool:\pset Ticket\\
bkd: Booked
\end{schema}
\nextside
\begin{schema}{KInit}
Kurbel
\ST
bkd= no
\end{schema}
\end{sidebyside}

\begin{sidebyside}
\begin{schema}{KBook}
\Delta Kurbel
\ST
bkd = no\\
kpool \neq \emptyset\\
bkd'= yes \\
kpool' = kpool
\end{schema}
\nextside
\begin{schema}{KArrive}
\Delta Kurbel\\
t!:Ticket
\ST
bkd = yes\\
kpool \neq \emptyset\\
bkd' = no\\
t!\mem kpool\\
kpool'=kpool\setminus \{t!\}
\end{schema}
\end{sidebyside}
}

The Marlowe specification is a downward simulation of the Kurbel (and in fact Kurbel is an upward simulation of Marlowe). The retrieve relation linking the two that one is tempted to write down is the following:

{\small
\begin{schema}{R}
Marlowe\\
Kurbel
\ST
bkd = no  \implies tkt = null \land kpool = mpool\\
bkd = yes \implies tkt \neq null \land 
   kpool = (mpool \union \{ ticket^{-1}(tkt) \})
\end{schema}
}
In \cite{SmithD06} a hand translation of these specifications into SAL was performed, followed by a merging into a single SAL specification - also performed by hand. A natural question to ask therefore is to what extent our automatic translation and combination is comparable with the manual process. The above candidate retrieve relation was used in the manual process, which revealed a failure to pass the necessary refinement conditions - both specification and retrieve relation needing adjustment before the Marlowe was shown to be a valid downward simulation of the Kurbel.

It is interesting to note that the results of the automatic translation were broadly comparable to the manual one, and in fact due to our optimizations show slight reduction in state space size (see table below). 
The automatic combination essentially identical to the manual. The latter is to be expected - the combination is essentially simple once the specifications have been converted into SAL.\\

\begin{tabular}{|c|c|c|} \hline
Step  & Manual & Auto \\ \hline
0    & 1344  &     840  \\
1     & 3360 &     6072  \\
2     & 8544 &     6072  \\
3     & 8544 &     6072 \\
4     & 8544 &     6072 \\ \hline
\end{tabular}

\section{Conclusion}
\label{sec:conclusions}

This work contributes on one hand to the strand of work on providing tool support for Z, and on the other hand to automatic refinement checking.

Recent work on providing tool support for Z includes the CZT (Community Z Tools) project \cite{Miller05a}, our own work 
\cite{DerrickNS06}, as well as related work such as ProZ \cite{PlLe07218}, 
which adapts the ProB \cite{Leuschel05a} tool for the Z notation.

Work on automatic refinement checking includes that of 
Bolton who has used Alloy to verify data refinements in Z 
\cite{Bolton05a}.
There have also been a
number of encoding of subsets of Z-based languages in the CSP model
checker FDR \cite{fis99a,mot01a,kas01a}, which checks that refinement holds between two
specifications by comparing the failures/divergences
semantics of the specifications; and simulation-based refinement
can be encoded as a failures/divergences check
\cite{Derrick02c,jos88a,he89a}.

Clearly there is much to be done in terms of further work here, not least some performance characterisations of when such an approach produces feasible state spaces.
\\~\\
{\bf Acknowledgements:} This work was done as part of collaborative work with Graeme Smith and Luke Wildman of
the University of Queensland. 
Tim Miller also gave valuable advice on the current CZT tools. 

\bibliographystyle{eptcs}

\bibliography{refine} 


\section*{Appendix A}
Here is the SAL translation of the concrete specification from Section \ref{sec:refinement}
{\small 
\begin{verbatim}
c : CONTEXT = BEGIN
NAT : TYPE = [0..4];
T : TYPE = {T__1, T__2, T__3, T__B};

State : MODULE =
BEGIN
 LOCAL max : NAT
 LOCAL l : sequence {T; T__B, 3} ! Sequence
 INPUT p? : T
 LOCAL invariant__ : BOOLEAN
 DEFINITION
   invariant__ = (sequence {T; T__B, 3} ! injective?(l) AND
     sequence {T; T__B, 3} ! valid?(l) AND
     p? /= T__B AND
     sequence {T; T__B, 3} ! size?(l) <= max)
 INITIALIZATION [
     l = sequence {T; T__B, 3} ! empty AND invariant__
   -->                                                                
 ]
 TRANSITION [
   CEnter :
       sequence {T; T__B, 3} ! size?(l) < max AND
       NOT set {T;} ! contains?(sequence {T; T__B, 3} ! range(l), p?) AND
       l' = sequence {T; T__B, 3} ! append(l, p?) AND
       invariant__ AND
       invariant__'
     -->
       l' IN {x : sequence {T; T__B, 3} ! Sequence | TRUE}
   []                                                                        
   CLeave :
       set {T;} ! contains?(sequence {T; T__B, 3} ! range(l), p?) AND
       l' = sequence {T; T__B, 3} ! remove(l,p?) AND
       invariant__ AND
       invariant__'
     -->
       l' IN {x : sequence {T; T__B, 3} ! Sequence | TRUE}
   []
   ELSE -->    l' = l                                                                  
 ]
END;
END
\end{verbatim}

\section*{Appendix B}
The result of automatically combining the two SAL modules from Z specifications given in Section \ref{sec:refinement}:
\begin{verbatim}
r2corr : CONTEXT = BEGIN

NAT : TYPE = [0..5];
T : TYPE = {T__1, T__2, T__3, T__B};
EVENT__ : TYPE = {AEnter, ALeave, CEnter, CLeave, Choose__};
T__counter : CONTEXT = count4 {T; T__1, T__2, T__3, T__B};

State : MODULE =
BEGIN
 LOCAL max : NAT
 LOCAL max : NAT
 LOCAL s : set {T;} ! Set
 INPUT p? : T
 LOCAL l : sequence {T; T__B, 3} ! Sequence
 LOCAL ev__ : EVENT__
 LOCAL invariant__ : BOOLEAN
 DEFINITION                                                                
   invariant__ = 
     (T__counter ! size?(s) <= max AND
     sequence {T; T__B, 3} ! injective?(l) AND
     p? /= T__B AND
     sequence {T; T__B, 3} ! valid?(l) AND
     sequence {T; T__B, 3} ! size?(l) <= max)
 INITIALIZATION [
     (s = sequence {T; T__B, 3} ! range(l))
   -->
 ]
 TRANSITION [
   AEnter :
       T__counter ! size?(s) < max AND                         
       NOT set {T;} ! contains?(s, p?) AND
       s' = set {T;} ! insert(s, p?) AND
       ev__' = AEnter AND
       invariant__ AND
       invariant__'
     -->
       s' IN {x : set {T;} ! Set | TRUE};
       l' IN {x : sequence {T; T__B, 3} ! Sequence | TRUE};
       ev__' IN {x : EVENT__ | TRUE}
   []
   ALeave :                                                                   
       set {T;} ! contains?(s, p?) AND
       s' = set {T;} ! remove(s, p?) AND
       ev__' = ALeave AND
       invariant__ AND
       invariant__'
     -->
       s' IN {x : set {T;} ! Set | TRUE};
       l' IN {x : sequence {T; T__B, 3} ! Sequence | TRUE};
       ev__' IN {x : EVENT__ | TRUE}
   []
   CEnter :                                                                   
       sequence {T; T__B, 3} ! size?(l) < max AND
       NOT set {T;} ! contains?(sequence {T; T__B, 3} ! range(l), p?) AND
       l' = sequence {T; T__B, 3} ! append(l, p?) AND
       ev__' = CEnter AND
       invariant__ AND
       invariant__'
     -->
       s' IN {x : set {T;} ! Set | TRUE};
       l' IN {x : sequence {T; T__B, 3} ! Sequence | TRUE};
       ev__' IN {x : EVENT__ | TRUE}
   []                                                                         
   CLeave :
       set {T;} ! contains?(sequence {T; T__B, 3} ! range(l), p?) AND
       l' = sequence {T; T__B, 3} ! remove(l,p?) AND
       ev__' = CLeave AND
       invariant__ AND
       invariant__'
     -->
       s' IN {x : set {T;} ! Set | TRUE};
       l' IN {x : sequence {T; T__B, 3} ! Sequence | TRUE};
       ev__' IN {x : EVENT__ | TRUE}                      
   []
   Choose__ :
       ev__' = Choose__ AND
       invariant__ AND
       invariant__'
     -->
       s' IN {x : set {T;} ! Set | TRUE};
       l' IN {x : sequence {T; T__B, 3} ! Sequence | TRUE};
       ev__' IN {x : EVENT__ | TRUE}
 ]
END;
END
\end{verbatim}

\end{document}